\documentclass[conference]{IEEEtran}
\pdfoutput=1

\IEEEoverridecommandlockouts                             
\usepackage{graphicx} 
\usepackage{amsmath} 
\usepackage{amssymb}  
\usepackage{amsfonts}                            
\usepackage{cite}
\usepackage{subfigure}                  
\usepackage{url}    
\usepackage[american,ngerman]{babel}
\usepackage{bm}

\usepackage{amsfonts}
\usepackage{bbm}
\usepackage{mathrsfs}
\usepackage{xspace}
\usepackage{bm}

\newenvironment{textbmatrix}{	\setlength{\arraycolsep}{2.5pt}%
								\big[\begin{matrix}}{\end{matrix}\big]%
								\raisebox{0.08ex}{\vphantom{M}}}

\def\be{\begin{equation}}
\def\ee{\end{equation}}
\def\een{\nonumber \end{equation}}
\def\mat{\begin{bmatrix}}
\def\emat{\end{bmatrix}}
\def\btm{\begin{textbmatrix}}
\def\etm{\end{textbmatrix}}

\def\ba#1\ea{\begin{align}#1\end{align}}
\def\bs#1\es{\begin{split}#1\end{split}} 
\def\bg#1\eg{\begin{gather}#1\end{gather}} 
\def\bi#1\ei{\begin{itemize}#1\end{itemize}}

\newcommand{\safemath}[2]{\newcommand{#1}{\ensuremath{#2}\xspace}}

\newcommand{\lefto}{\mathopen{}\left}

\DeclareMathOperator{\sign}{sign}			
\DeclareMathOperator{\Exop}{\mathbb{E}}		

\newcommand{\Ex}[1]{\ensuremath{\Exop\lefto[#1\right]}} 	
\safemath{\normal}{\mathcal{N}}				
\safemath{\complexnormal}{\mathcal{CN}}				

\safemath{\circnorm}{\mathcal{CN}}			

\safemath{\interior}{\mathrm{Int}}			

\safemath{\dfn}{:=}							
\safemath{\markov}{\leftrightarrow}			

\safemath{\SNR}{\text{\sc snr}} 				
\safemath{\No}{N_0}							
\safemath{\Eb}{E_b}							
\safemath{\EbNo}{\frac{\Eb}{\No}}
\safemath{\EsNo}{\frac{\Es}{\No}}

\DeclareMathOperator{\CHop}{\ensuremath{\mathbb{H}}} 
\safemath{\tvir}{h_{\CHop}}					
\safemath{\tvtf}{L_{\CHop}}					
\safemath{\spf}{S_{\CHop}}						
\safemath{\bff}{H_{\CHop}}					

\safemath{\ircf}{R_{h}}						
\safemath{\scf}{R_{S}}						
	\safemath{\CH}{C_{\CHop}}
\safemath{\tvcf}{R_{L}}						
\safemath{\bfcf}{R_{H}}						
\safemath{\RH}{R_{\CHop}}

\safemath{\mi}{I}							
\safemath{\capacity}{C}						

\safemath{\dB}{\,\mathrm{dB}}
\safemath{\dBm}{\,\mathrm{dBm}}
\safemath{\Hz}{\,\mathrm{Hz}}
\safemath{\kHz}{\,\mathrm{kHz}}
\safemath{\MHz}{\,\mathrm{MHz}}
\safemath{\GHz}{\,\mathrm{GHz}}
\safemath{\s}{\,\mathrm{s}}
\safemath{\ms}{\,\mathrm{ms}}
\safemath{\mus}{\,\mathrm{\mu s}}
\safemath{\ns}{\,\mathrm{ns}}
\safemath{\meter}{\,\mathrm{m}}
\safemath{\mm}{\,\mathrm{mm}}
\safemath{\cm}{\,\mathrm{cm}}
\safemath{\km}{\, \mathrm{km}}
\safemath{\m}{\,\mathrm{m}}
\safemath{\J}{\,\mathrm{J}}
\safemath{\K}{\,\mathrm{K}}
\safemath{\bit}{\,\mathrm{bit}}

\safemath{\define}{\triangleq}			

\safemath{\equivalent}{\sim}
\safemath{\distas}{\sim}					

\safemath{\reals}{\mathbb{R}}
\safemath{\positivereals}{\mathbb{R}^{+}}
\safemath{\integers}{\mathbb{Z}}
\safemath{\posint}{\mathbb{Z}_{+}}
\safemath{\naturals}{\mathbb{N}}
\safemath{\complexset}{\mathbb{C}}

\safemath{\setA}{\mathcal{A}}
\safemath{\setB}{\mathcal{B}}
\safemath{\setC}{\mathcal{C}}
\safemath{\setD}{\mathcal{D}}
\safemath{\setE}{\mathcal{E}}
\safemath{\setF}{\mathcal{F}}
\safemath{\setG}{\mathcal{G}}
\safemath{\setH}{\mathcal{H}}
\safemath{\setI}{\mathcal{I}}
\safemath{\setJ}{\mathcal{J}}
\safemath{\setK}{\mathcal{K}}
\safemath{\setL}{\mathcal{L}}
\safemath{\setM}{\mathcal{M}}
\safemath{\setN}{\mathcal{N}}
\safemath{\setO}{\mathcal{O}}
\safemath{\setP}{\mathcal{P}}
\safemath{\setQ}{\mathcal{Q}}
\safemath{\setR}{\mathcal{R}}
\safemath{\setS}{\mathcal{S}}
\safemath{\setT}{\mathcal{T}}
\safemath{\setU}{\mathcal{U}}
\safemath{\setV}{\mathcal{V}}
\safemath{\setW}{\mathcal{W}}
\safemath{\setX}{\mathcal{X}}
\safemath{\setY}{\mathcal{Y}}
\safemath{\setZ}{\mathcal{Z}}
\safemath{\emptySet}{\varnothing}

\safemath{\bma}{\mathbf{a}}
\safemath{\bmb}{\mathbf{b}}
\safemath{\bmc}{\mathbf{c}}
\safemath{\bmd}{\mathbf{d}}
\safemath{\bme}{\mathbf{e}}
\safemath{\bmf}{\mathbf{f}}
\safemath{\bmg}{\mathbf{g}}
\safemath{\bmh}{\mathbf{h}}
\safemath{\bmi}{\mathbf{i}}
\safemath{\bmj}{\mathbf{j}}
\safemath{\bmk}{\mathbf{k}}
\safemath{\bml}{\mathbf{l}}
\safemath{\bmm}{\mathbf{m}}
\safemath{\bmn}{\mathbf{n}}
\safemath{\bmo}{\mathbf{o}}
\safemath{\bmp}{\mathbf{p}}
\safemath{\bmq}{\mathbf{q}}
\safemath{\bmr}{\mathbf{r}}
\safemath{\bms}{\mathbf{s}}
\safemath{\bmt}{\mathbf{t}}
\safemath{\bmu}{\mathbf{u}}
\safemath{\bmv}{\mathbf{v}}
\safemath{\bmx}{\mathbf{x}}
\safemath{\bmy}{\mathbf{y}}
\safemath{\bmz}{\mathbf{z}}

\bmdefine{\biad}{a}
\bmdefine{\bibd}{b}
\bmdefine{\bicd}{c}
\bmdefine{\bidd}{d}
\bmdefine{\bied}{e}
\bmdefine{\bifd}{f}
\bmdefine{\bigd}{g}
\bmdefine{\bihd}{h}
\bmdefine{\biid}{i}
\bmdefine{\bijd}{j}
\bmdefine{\bikd}{k}
\bmdefine{\bild}{l}
\bmdefine{\bimd}{m}
\bmdefine{\bind}{n}
\bmdefine{\biod}{o}
\bmdefine{\bipd}{p}
\bmdefine{\biqd}{q}
\bmdefine{\bird}{r}
\bmdefine{\bisd}{s}
\bmdefine{\bitd}{t}
\bmdefine{\biud}{u}
\bmdefine{\bivd}{v}
\bmdefine{\biwd}{w}
\bmdefine{\bixd}{x}
\bmdefine{\biyd}{y}
\bmdefine{\bizd}{z}

\bmdefine{\bixid}{\xi}
\bmdefine{\bilambdad}{\lambda}
\bmdefine{\bimud}{\mu}
\bmdefine{\bithetad}{\theta}
\bmdefine{\biphid}{\phi}
\bmdefine{\bipi}{\pi}

\safemath{\bmia}{\biad}
\safemath{\bmib}{\bibd}
\safemath{\bmic}{\bicd}
\safemath{\bmid}{\bidd}
\safemath{\bmie}{\bied}
\safemath{\bmif}{\bifd}
\safemath{\bmig}{\bigd}
\safemath{\bmih}{\bihd}
\safemath{\bmii}{\biid}
\safemath{\bmij}{\bijd}
\safemath{\bmik}{\bikd}
\safemath{\bmil}{\bild}
\safemath{\bmim}{\bimd}
\safemath{\bmin}{\bind}
\safemath{\bmio}{\biod}
\safemath{\bmip}{\bipd}
\safemath{\bmiq}{\biqd}
\safemath{\bmir}{\bird}
\safemath{\bmis}{\bisd}
\safemath{\bmit}{\bitd}
\safemath{\bmiu}{\biud}
\safemath{\bmiv}{\bivd}
\safemath{\bmiw}{\biwd}
\safemath{\bmix}{\bixd}
\safemath{\bmiy}{\biyd}
\safemath{\bmiz}{\bizd}

\safemath{\bmxi}{\bixid}
\safemath{\bmlambda}{\bilambdad}
\safemath{\bmmu}{\bimud}
\safemath{\bmtheta}{\bithetad}
\safemath{\bmphi}{\biphid}
\safemath{\bmpi}{\bipi}

\safemath{\bA}{\mathbf{A}}
\safemath{\bB}{\mathbf{B}}
\safemath{\bC}{\mathbf{C}}
\safemath{\bD}{\mathbf{D}}
\safemath{\bE}{\mathbf{E}}
\safemath{\bF}{\mathbf{F}}
\safemath{\bG}{\mathbf{G}}
\safemath{\bH}{\mathbf{H}}
\safemath{\bI}{\mathbf{I}}
\safemath{\bJ}{\mathbf{J}}
\safemath{\bK}{\mathbf{K}}
\safemath{\bL}{\mathbf{L}}
\safemath{\bM}{\mathbf{M}}
\safemath{\bN}{\mathbf{N}}
\safemath{\bO}{\mathbf{O}}
\safemath{\bP}{\mathbf{P}}
\safemath{\bQ}{\mathbf{Q}}
\safemath{\bR}{\mathbf{R}}
\safemath{\bS}{\mathbf{S}}
\safemath{\bT}{\mathbf{T}}
\safemath{\bU}{\mathbf{U}}
\safemath{\bV}{\mathbf{V}}
\safemath{\bW}{\mathbf{W}}
\safemath{\bX}{\mathbf{X}}
\safemath{\bY}{\mathbf{Y}}
\safemath{\bZ}{\mathbf{Z}}

\safemath{\bZero}{\mathbf{0}}

\bmdefine{\biAd}{A}
\bmdefine{\biBd}{B}
\bmdefine{\biCd}{C}
\bmdefine{\biDd}{D}
\bmdefine{\biEd}{E}
\bmdefine{\biFd}{F}
\bmdefine{\biGd}{G}
\bmdefine{\biHd}{H}
\bmdefine{\biId}{I}
\bmdefine{\biJd}{J}
\bmdefine{\biKd}{K}
\bmdefine{\biLd}{L}
\bmdefine{\biMd}{M}
\bmdefine{\biOd}{N}
\bmdefine{\biPd}{O}
\bmdefine{\biQd}{P}
\bmdefine{\biRd}{R}
\bmdefine{\biSd}{S}
\bmdefine{\biTd}{T}
\bmdefine{\biUd}{U}
\bmdefine{\biVd}{V}
\bmdefine{\biWd}{W}
\bmdefine{\biXd}{X}
\bmdefine{\biYd}{Y}
\bmdefine{\biZd}{Z}

\bmdefine{\biDelta}{\Delta}
\bmdefine{\biLambda}{\Lambda}
\bmdefine{\biPhi}{\Phi}
\bmdefine{\biSigma}{\Sigma}
\bmdefine{\biOmega}{\Omega}
\bmdefine{\biTheta}{\Theta}

\safemath{\bimA}{\biAd}
\safemath{\bimB}{\biBd}
\safemath{\bimC}{\biCd}
\safemath{\bimD}{\biDd}
\safemath{\bimE}{\biEd}
\safemath{\bimF}{\biFd}
\safemath{\bimG}{\biGd}
\safemath{\bimH}{\biHd}
\safemath{\bimI}{\biId}
\safemath{\bimJ}{\biJd}
\safemath{\bimK}{\biKd}
\safemath{\bimL}{\biLd}
\safemath{\bimM}{\biMd}
\safemath{\bimN}{\biNd}
\safemath{\bimO}{\biOd}
\safemath{\bimP}{\biPd}
\safemath{\bimQ}{\biQd}
\safemath{\bimR}{\biRd}
\safemath{\bimS}{\biSd}
\safemath{\bimT}{\biTd}
\safemath{\bimU}{\biUd}
\safemath{\bimV}{\biVd}
\safemath{\bimW}{\biWd}
\safemath{\bimX}{\biXd}
\safemath{\bimY}{\biYd}
\safemath{\bimZ}{\biZd}

\safemath{\bDelta}{\bielta}
\safemath{\bLambda}{\biLambda}
\safemath{\bPhi}{\biPhi}
\safemath{\bSigma}{\biSigma}
\safemath{\bOmega}{\biOmega}
\safemath{\bTheta}{\biTheta}

\safemath{\veca}{\bma}
\safemath{\vecb}{\bmb}
\safemath{\vecc}{\bmc}
\safemath{\vecd}{\bmd}
\safemath{\vece}{\bme}
\safemath{\vecf}{\bmf}
\safemath{\vecg}{\bmg}
\safemath{\vech}{\bmh}
\safemath{\veci}{\bmi}
\safemath{\vecj}{\bmj}
\safemath{\veck}{\bmk}
\safemath{\vecl}{\bml}
\safemath{\vecm}{\bmm}
\safemath{\vecn}{\bmn}
\safemath{\veco}{\bmo}
\safemath{\vecp}{\bmp}
\safemath{\vecq}{\bmq}
\safemath{\vecr}{\bmr}
\safemath{\vecs}{\bms}
\safemath{\vect}{\bmt}
\safemath{\vecu}{\bmu}
\safemath{\vecv}{\bmv}
\safemath{\vecw}{\bmw}
\safemath{\vecx}{\bmx}
\safemath{\vecy}{\bmy}
\safemath{\vecz}{\bmz}
\safemath{\vecZero}{\bZero}

\safemath{\vecxi}{\bmxi}
\safemath{\veclambda}{\bmlambda}
\safemath{\vecmu}{\bmmu}
\safemath{\vectheta}{\bmtheta}
\safemath{\vecphi}{\bmphi}
\safemath{\vecpi}{\bmpi}

\safemath{\matA}{\bA}
\safemath{\matB}{\bB}
\safemath{\matC}{\bC}
\safemath{\matD}{\bD}
\safemath{\matE}{\bE}
\safemath{\matF}{\bF}
\safemath{\matG}{\bG}
\safemath{\matH}{\bH}
\safemath{\matI}{\bI}
\safemath{\matJ}{\bJ}
\safemath{\matK}{\bK}
\safemath{\matL}{\bL}
\safemath{\matM}{\bM}
\safemath{\matN}{\bN}
\safemath{\matO}{\bO}
\safemath{\matP}{\bP}
\safemath{\matQ}{\bQ}
\safemath{\matR}{\bR}
\safemath{\matS}{\bS}
\safemath{\matT}{\bT}
\safemath{\matU}{\bU}
\safemath{\matV}{\bV}
\safemath{\matW}{\bW}
\safemath{\matX}{\bX}
\safemath{\matY}{\bY}
\safemath{\matZ}{\bZ}
\safemath{\matZero}{\bZero}

\safemath{\matDelta}{\bDelta}
\safemath{\matLambda}{\bLambda}
\safemath{\matPhi}{\bPhi}
\safemath{\matSigma}{\bSigma}
\safemath{\matOmega}{\bOmega}
\safemath{\matTheta}{\bTheta}

\newtheorem{theorem}{Theorem}

\bmdefine{\balpha}{\alpha}

\title{\LARGE \bf Distributed Spatial Multiplexing with $\mathbf{1}$-Bit Feedback}
\author{Jatin Thukral and Helmut B\"olcskei
\thanks{This research was supported in part by the Swiss National Science Foundation (SNF) under grant No. 200020-109619 and by the Nokia Research Center Helsinki, Finland.}
\thanks{The authors are with ETH Zurich, Switzerland, (Email: {\tt\small \{jatin,boelcskei\}@nari.ee.ethz.ch})}}

\begin{document}

\maketitle
\begin{abstract}
We analyze the feasibility of distributed spatial multiplexing with limited feedback in a slow-fading interference network with $MN$ non-cooperating single-antenna sources and $M$ non-cooperating single-antenna destinations. In particular, we assume that the sources are divided into $M$ mutually exclusive groups of $N$ sources each, every group is dedicated to transmit a common message to a unique destination, all transmissions occur concurrently and in the same frequency band and a dedicated $1$-bit  broadcast feedback channel from each destination to its corresponding group of sources exists. We provide a feedback-based iterative distributed (multi-user) beamforming algorithm, which ``learns'' the channels between each group of sources and its assigned destination. This algorithm is a straightforward generalization, to the multi-user case, of the feedback-based iterative distributed beamforming algorithm proposed recently by Mudumbai, Hespanha, Madhow and Barriac in \cite{Mudumbai2006Distributed-Tra} for networks with a single group of sources and a single destination. Putting the algorithm into a Markov chain context, we provide a simple convergence proof. We then show that, for $M$ finite and $N\,\rightarrow\,\infty$, spatial multiplexing based on the beamforming weights produced by the algorithm achieves full spatial multiplexing gain of $M$ and full per-stream array gain of $N$, provided the time spent ``learning'' the channels scales linearly in $N$. The 
network is furthermore shown to ``crystallize'' in the sense that, in the large-$N$ limit, the $M$ individual fading links not only decouple (as reflected by full spatial multiplexing gain) but also converge to non-fading links. Finally, we quantify the impact of the performance of the iterative distributed beamforming algorithm on the crystallization rate, and we show that the multi-user nature of the network leads to a significant reduction in the crystallization rate, when compared to the $M=1$ case.
\end{abstract}

\section{INTRODUCTION}
\label{SISOAllertonIntroduction}

We consider a special class of interference networks, where $MN$ non-cooperating sources are divided into $M$ mutually exclusive groups $\mathcal{G}_i, i=1,2,\ldots,M$, such that the $N$ sources in the $i_{th}$ group, 
denoted as $\mathcal{S}_i^j$, $j=1,2,\ldots ,N$, are dedicated to transmit, through slow-fading channels, a common message to their assigned single-antenna destination $\mathcal{D}_i$.
All transmissions occur concurrently and in the same frequency band and the destinations $\mathcal{D}_{i}$ do not cooperate.
This network models the second hop of the coherent multi-user relaying protocol in \cite{Bolcskei2006Capacity-scalin} under the assumption that the first hop transmission is error-free. The results in \cite{Bolcskei2006Capacity-scalin} imply that, for the interference network considered in this paper, for $M$ fixed and $N\rightarrow\infty$,  full spatial multiplexing gain of $M$ and a per-stream (distributed) array gain of $N$ 
can be obtained, provided that each source knows the channel to its assigned destination perfectly. In this paper, we analyze the case where the perfect channel state information assumption is relaxed to having a $1$-bit broadcast feedback channel from each destination $\mathcal{D}_i$ to its sources $\mathcal{G}_i$. These broadcast feedback channels are non-interfering. We provide a feedback-based iterative distributed (multi-user) beamforming algorithm, which ``learns'' the channels between each group of sources and its assigned destination. This algorithm is a straightforward generalization, to the multi-user case, of the feedback-based iterative distributed beamforming algorithm proposed recently by Mudumbai, Hespanha, Madhow and Barriac in \cite{Mudumbai2006Distributed-Tra} for networks with a single group of sources and a single destination. Making the simplifying assumption, compared to \cite{Mudumbai2006Distributed-Tra}, of the fading coefficients as well as all the signals being real-valued allows us to put the iterative algorithm into a Markov chain context, thereby setting the stage for a simple convergence proof. We then show that, for $M$ finite and $N\,\rightarrow\,\infty$,
spatial multiplexing based on the beamforming weights produced by the iterative algorithm achieves full spatial multiplexing gain of $M$ and full per-stream  (distributed) array gain of $N$, provided the time spent ``learning'' the channels scales linearly in $N$. We furthermore demonstrate that the $M$ effective links $\mathcal{G}_{i} \rightarrow \mathcal{D}_{i}$ in the network not only decouple (reflected by full spatial multiplexing gain) but also converge to non-fading links as $N\rightarrow\infty$, i.e., in the terminology of \cite{Morgenshtern2006Crystallization}, the network ``crystallizes''. Finally, we quantify the impact of the performance of the iterative algorithm on the crystallization rate, and we show that the multi-user nature of the network leads to a significant reduction in the crystallization rate, when compared to the $M=1$ case. 

{\it Notation}\/: The superscripts $^T$ and $^{-1}$ stand for transposition and inverse, respectively. $|\mathcal{G}|$ denotes  the cardinality of the set $\mathcal{G}$, $|x|$ is the absolute value of the scalar $x$, 
and $\lfloor a \rfloor$ denotes the greatest integer that is smaller than or equal to the real number $a$. $\mathcal{N}(\mu,\sigma^2)$ stands for the normal distribution with mean $\mu$ and variance $\sigma^2$. $\log(\cdot)$ denotes 
logarithm to the base 2. $f_X(\cdot)$ stands for the probability density function (p.d.f.) of the random variable $X$ and $X\sim Y$ denotes equivalence in distribution. $\mathcal{A}\equiv\mathcal{B}$ denotes that the sets (of terminals) $\mathcal{A}$ and $\mathcal{B}$ are equal. $P(\omega)$ is the probability of event $\omega$, $\mathbb{E}[X]$ and VAR$[X]$ are the expected value and the variance, respectively, of the random variable $X$ and w.p. stands for \emph{with probability}.
Since the terminals in $\mathcal{G}_i$ are assumed to have a common message for their assigned destination $\mathcal{D}_i$, we will be using the notation  $\mathcal{G}_i\rightarrow\mathcal{D}_i$ to denote the corresponding single-input single-output link between the group $\mathcal{G}_i$ and destination $\mathcal{D}_i$. Vectors and matrices are set in lower-case and upper-case bold-face letters, respectively. 

\section{System and Signal Model}
\label{sec:SingleAntennaSystemModel}
We assume that $x_{i}[n]$ is the common message of the sources in $\mathcal{G}_{i}$ to be transmitted to $\mathcal{D}_{i}$. In the remainder of the paper, we distinguish between a training phase during which the $N$ scalar channels between $\mathcal{S}_i^j, j=1,2,\ldots,N,$ and  $\mathcal{D}_i$ are ``learned'' for each $i\,\in\,\{1,2,\ldots,M\},$  and a data transmission phase following the training phase. During the training phase for $\mathcal{G}_i$ the feedback broadcast channel $\mathcal{D}_i\rightarrow\mathcal{G}_i$ is used once every frame of $T_f$ time slots. The $l_{th}$ frame ($l=0,1,...$), denoted by $\mathcal{F}_l$, consists of the time slots   $n=lT_f,lT_f+1,\ldots, (l+1)T_f-1$. In the $l_{th}$  frame, each source $\mathcal{S}_i^j$ multiplies the sequence $x_i[n]$ (which is a training sequence during the training phase) by a corresponding beamforming weight $\alpha_i^j[l]\in\{1,-1\}$ before transmission; these beamforming weights are kept constant during the entire frame. We furthermore assume that all the channels in the network are flat-fading and remain constant throughout the entire time-interval of interest, i.e., during training and data transmission phases. Throughout the paper, we assume that $M$ is finite. The corresponding input-output relations are now given by
\ba
y_i[n]&=\bigg(\sum_{j=1}^{N}h_{i,i}^j \alpha_i^j[l]\bigg)x_i[n]+\underbrace{\mathop{\sum_{r=1}^M}_{r\neq i} \bigg(\sum_{j=1}^N h_{i,r}^j\alpha_r^j[l]\bigg)x_r[n]}_{\text{interference}}\notag\\
\label{SISOInterferenceAllertonMainIORelation}
&\qquad+\ w_i[n],\qquad i=1,2,\ldots,M,
\ea
where $l=\left\lfloor \frac{n}{T_f}\right\rfloor $, $y_i[n]\in\mathbb{R}$ denotes the symbol received at $\mathcal{D}_i$ in the $n_{th}$ time slot, $h_{i,r}^j\in\mathbb{R}$ stands for the fading coefficient between $\mathcal{S}_r^j$ and $\mathcal{D}_i$ and $w_i[n]$ denotes the $\mathcal{N}(0,N_o)$ i.i.d. noise sequence at $\mathcal{D}_i$. The fading coefficients $h_{i,r}^j, \mbox{for all}\,\, i,r,j,$ are assumed i.i.d. $\normal{(0,1)}$. The signals $x_i[n]$ obey the average power constraint
\ba
\mathbb{E}\hspace{-0.8mm}\left[ |x_i[n]|^2\right] \leq \frac{P}{N} ,\qquad i=1,2,\ldots,M,\notag
\ea
so that the average power transmitted by each group $\mathcal{G}_i$ is limited by $P$. During the training phase, the frame-rate sequences $\alpha_i^j[l]$ are updated based on the $1$-bit feedback received at the end of each frame. The goal of this process is to find the beamforming weights $\alpha_i^j=\sign(h_{i,i}^j)$. While it is in general difficult to put the individual groups into perfect beamforming configuration, so that 
\ba
\sum_{j=1}^Nh_{i,i}^j\alpha_i^j = \sum_{j=1}^N|h_{i,i}^j|,\quad \text{for group $\mathcal{G}_i$},\notag
\ea
we will show that if in each group a sufficient number of sources is in beamforming configuration, full spatial multiplexing gain can be achieved. In what follows, we shall say that in a given group $\mathcal{G}_i$, sources satisfying $h_{i,i}^j\alpha_i^j>0$ are aligned, whereas sources with $h_{i,i}^j\alpha_i^j<0$ are reverse-aligned. We denote the final beamforming weights $\alpha_i^j$ produced during the training phase as $\tilde{\alpha}_i^j$. In the data transmission phase, the groups $\mathcal{G}_i, i=1,2,\ldots,M,$ perform co-channel data transmission (i.e., spatial multiplexing) based on $\tilde{\alpha}_i^j$ so that the corresponding input-output relation is given by \eqref{SISOInterferenceAllertonMainIORelation} with $\alpha_i^j[l]=\tilde{\alpha}_i^j$.


\section{Iterative Distributed Beamforming}
\label{sec:SingleAntennaSchemeDetails}
We shall next describe the algorithm carried out during the training phase. The overall training phase is assumed to consist of $T_{tr}=k_oMN$ frames (each of which contains $T_f$ time slots) divided into $M$  blocks of $k_oN$ frames each. The role of the parameter $k_o$ will become clear later. During each of these $M$ blocks, precisely one of the groups $\mathcal{G}_i$ follows the three-step iterative distributed beamforming algorithm, described below, while all the other groups of sources $\mathcal{G}_r,r\neq i,$ remain silent. At the end of the training phase of $k_oMN$ frames, each of the groups $\mathcal{G}_i$ is in (close-to) beamforming configuration with respect to (w.r.t.) its assigned destination. 
The order in which the groups follow the three-step procedure below can be decided offline and communicated to all the nodes in the network. Without loss of generality (w.l.o.g.), we assume that the group $\mathcal{G}_i$ is being processed during the $i_{th}$ block defined as the set of frames $l = (i-1)k_oN,(i-1)k_oN+1,\ldots ,ik_oN-1$. Since much of the analysis in the current and the next section deals with a single group $\mathcal{G}_i$ only, we shall consider, w.l.o.g., the group $\mathcal{G}_{1}$, drop the index $i=1$ and, 
wherever appropriate, use the convention $\mathcal{G}\equiv\mathcal{G}_1,\mathcal{S}^j\equiv\mathcal{S}_1^j , \mathcal{D}\equiv\mathcal{D}_1,\alpha^j=\alpha_1^j,\hat{\alpha}^j=\hat{\alpha}_1^j$ and $h^j=h_{1,1}^j$. The three steps carried out by the iterative distributed beamforming algorithm can now be summarized as follows. 
\begin{itemize}
\item{{\it Step 1. Initialization of the received signal level}:} This step pertains to the zeroth frame in the  block. Each of the sources $\mathcal{S}^j$ initializes its beamforming weight according to $\alpha^j[0]=1$, initializes an auxiliary beamforming weight as $\hat{\alpha}^j[0]=\alpha^j[0]$ and starts transmitting the pilot symbol \hspace{-0.6mm}$\sqrt{P/N}$. The corresponding received signal at destination $\mathcal{\mathcal{D}}$ is given by 
\ba
y[n]&=\sqrt{\frac{P}{N}}\sum_{j=1}^{N}h^j\alpha^j[0]+w[n]\notag\\
&=\sqrt{\frac{P}{N}}\sum_{j=1}^{N}h^j+w[n],\qquad n\in\mathcal{F}_{0}. \notag
\ea
The destination $\mathcal{D}$ then estimates the received signal level by averaging $y[n]$ over the entire frame, resulting in
\ba
L^{\text{rx}} 
&=\frac{1}{T_f}\sum_{n \in \mathcal{F}_{0}}\!\! y[n]=\sqrt{\frac{P}{N}} \sum_{j=1}^{N}h^j. \notag
\ea
Here, we assumed that the estimate of $L^{\text{rx}}$ is perfect, which requires that $T_f$ be sufficiently large. Finally, $\mathcal{D}$ initializes $L^{\text{max}}=L^{\text{rx}}$.
\item{{\it Step 2. Iterative distributed beamforming}:} This is the iterative step that is performed for each frame, except for the zeroth one, i.e., for $l=1,2,\ldots,k_oN-1$ (recall that 
the zeroth frame is used to carry out the previous initialization step). The details of this step are as follows. At the beginning of each frame, each of the sources $\mathcal{S}^j, j=1,2,\ldots,N,$ chooses its beamforming weight $\alpha^j[l]$, independently across $j$, according to:
\ba
\label{RandomPerturbation}
\alpha^j[l]&=
\begin{cases}  \hat{\alpha}^j[l-1], & w.p.\ \  1-\frac{1}{N} \\
                -\hat{\alpha}^j[l-1], & w.p.\ \  \frac{1}{N}
                \end{cases}.
\ea
Each of the sources $\mathcal{S}^j$ then transmits the pilot  symbol $\sqrt{P/N}$ throughout the frame,  using the beamforming weight $\alpha^j[l]$, i.e., $\mathcal{S}^j$ transmits the signal $\alpha^j[l]\sqrt{P/N}$. At the end of the frame under consideration, $\mathcal{D}$ estimates the corresponding received signal level, as in the initialization step, according to\footnote{Again, we assume the estimate to be perfect.}
\ba
\label{IterativePowerCalculation}
L^{\text{rx}} &=\frac{1}{T_f}\sum_{n\in\mathcal{F}_l}y[n] 
= \sqrt{\frac{P}{N}} \sum_{j=1}^{N}h^j\alpha^j[l]
\ea
and, through the 1-bit broadcast feedback channel, informs all the sources in $\mathcal{G}$ whether 
$L^{\text{rx}} > L^{\text{max}}$ or not. Based on the received feedback, the sources $\mathcal{S}^j, j=1,2,\ldots,N,$ update their auxiliary beamforming weights $\hat{\alpha}^j[l]$ as follows:
\ba
\label{AuxiliaryParameterUpdateRule}
\hat{\alpha}^j[l]&=
\begin{cases}  
        \hat{\alpha}^j[l-1], & \text{if } L^{\text{rx}} \leq L^{\text{max}} \\
        \alpha^j[l], & \text{if } L^{\text{rx}} > L^{\text{max}}
\end{cases},\  j=1,2,\ldots,N.
\ea
Finally, if $L^{\text{rx}} > L^{\text{max}}$, $\mathcal{D}$ updates 
$L^{\text{max}}=L^{\text{rx}}$.

\item{{\it Step 3. Silencing}:} At the end of the block, the sources in $\mathcal{G}$ store the current values of their respective auxiliary beamforming weights as $\tilde{\alpha}^j=\hat{\alpha}^j[k_oN-1]$ and go silent.
\end{itemize}
%

In the proposed protocol, only one group of sources is active during a given block. It is therefore natural to ask whether the groups that have finished their training phase could start their data transmission phase while the remaining groups are ``learning'' their channels through the three-step procedure above. At first sight, one would be tempted to conclude that such a modified protocol would result in higher spectral efficiency. We note, however, that performing Steps 1 and 2 above in the presence of interference created by the groups already transmitting data would require a longer $T_f$, say $T_f^I$, in order to achieve the same quality of the received signal level estimate as in the original protocol, i.e., in the absence of interference. The resulting tradeoff can be illustrated roughly by assuming \emph{maximum-likelihood} estimation of the parameter $L^{\text{rx}}$. The corresponding variance of the estimation error is given by $\sigma^2=\frac{N_o}{T_f}$ in the original protocol and by $\sigma_M^2=\frac{(N_o+\sigma_I^2)}{T_f^I}$ in the modified protocol, where, considering the link $\mathcal{G}_{i}\,\rightarrow\,\mathcal{D}_{i}$, we have 
\vspace*{-1mm}
\ba
\label{ProtocolComparisonInterferencePower}
\sigma_I^2&=\frac{P}{N}\sum_{r=1}^{i-1}\left|\sum_{j=1}^Nh_{i,r}^j\tilde{\alpha}_r^j\right|^2.
\ea
Consequently, if we require the variance of the estimation error in the two protocols to be equal, then
\ba
\label{ProtocolTradeoffFrameRatio}
\frac{T_f^I}{T_f}&=1+\frac{\sigma_I^2}{N_o}.
\ea
Assuming that in each of the links already in (close-to) beamforming configuration the corresponding group of sources transmits at a  rate of $R$ bits per time slot, the total number of bits transmitted in the modified protocol during the first $k_oMNT_f^I$ time slots is given by $Rk_0NT_f^I M(M-1)/ 2$ [bit] whereas it is $Rk_oM^2N(T_f^I-T_f)$ [bit] in the original protocol. Using \eqref{ProtocolTradeoffFrameRatio}, this implies that the modified protocol would have a higher spectral efficiency if 
\ba
\label{ConditionProtocolOneBetter}
\frac{\sigma_I^2}{N_o}&< 1-\frac{2}{M+1}. 
\ea
Now, since the weights $\tilde{\alpha}_i^j$ have been optimized to be in beamforming configuration w.r.t. $h_{i,i}^{j}$, they are independent of $h_{i,r}^j$ for $r\,\neq\,i$.
Furthermore, $\tilde{\alpha}_i^j=\sign(h_{i,i}^j)$ and $\tilde{\alpha}_i^j =-\sign(h_{i,i}^j)$ for $\mathcal{S}_i^j$ in the set of aligned and reverse-aligned sources, respectively. Since $P(\sign(h_{i,i}^j)=1)=P(\sign(h_{i,i}^j)=-1)=1/2$, it follows that $P(\tilde{\alpha}_i^j=1)=P(\tilde{\alpha}_i^j=-1)=1/2$, for all $j$. Along with the assumption  $h_{i,r}^j\sim\mathcal{N}(0,1)$, for all $i,r,j$, we therefore get $h_{i,r}^j\tilde{\alpha}_r^j\sim h_{i,r}^j,$ for all $j$ and $\,r\,<\,i$, which allows us to conclude that $\left|\sum_{j=1}^Nh_{i,r}^j\tilde{\alpha}_r^j\right|^2, r\,<\,i$, scales linearly in $N$. Consequently, it follows from \eqref{ProtocolComparisonInterferencePower} that $\sigma_I^2$ is proportional to $P$, which implies that the condition \eqref{ConditionProtocolOneBetter} would not be met for any reasonable choice of 
SNR $=P/N_o$. We can therefore conclude that, provided the variance of the signal level estimation error is the relevant performance measure, the original protocol would in practice always outperform the modified protocol in terms of spectral efficiency.

\section{Convergence of The Iterative Distributed Beamforming Algorithm}
We shall next show that the iterative algorithm described in the previous section converges to the beamforming configuration for $k_o$ large. The proof is rather straightforward and consists of the following sequence of arguments:
\begin{itemize}
\item{} We start by recognizing that through \eqref{RandomPerturbation} and \eqref{AuxiliaryParameterUpdateRule}, the coefficient vector $\hat{\balpha}[l]=[\hat{\alpha}^1[l]\ \hat{\alpha}^2[l]\ \ldots\ \hat{\alpha}^N[l] ]^T$ depends only on the coefficient vector  in the previous frame, i.e., on $\hat{\balpha}[l-1]$. The sequence of vectors $\hat{\balpha}[l]$ therefore follows a Markov chain.
\item{} Each of the elements in the vector $\hat{\balpha}[l]$ can take on the values $1$ or $-1$, which implies that the total number of states of the Markov chain is given by $2^N$. Let us denote these states by $\mathcal{Q}_1,\mathcal{Q}_2,\cdots,\mathcal{Q}_{2^N}$. To each of the states, we can associate a value $\sum_{j=1}^N h^j\hat{\alpha}^j$. 
\item{} The Markov chain has one special state, namely, when $\hat{\alpha}^j=\mathrm{sign}(h^j)$, for all  $j$.  W.l.o.g. we let this state be $\mathcal{Q}_1$ and note that it corresponds to the beamforming configuration, i.e.,  
\ba
\sum_{j=1}^Nh^j\hat{\alpha}^j = \sum_{j=1}^Nh^j \mathrm{sign}(h^j)=\sum_{j=1}^N|h^j|
\ea
which implies that once the system is in state $\mathcal{Q}_1$ the parameter  $L^{\text{max}}$ is maximized over all states $\mathcal{Q}_i,i=1,2,\ldots,2^N$. From the update rule \eqref{AuxiliaryParameterUpdateRule}, we can therefore conclude that once in $\mathcal{Q}_1$ the  Markov chain will remain in $\mathcal{Q}_1$, which implies that $\mathcal{Q}_1$ is an \emph{absorbing state} (see\cite[Def. 11.1]{Grinstead1997Introduction-to}).
\item{} Each of the states $\mathcal{Q}_r, r\neq 1,$ corresponds to a coefficient vector $\hat{\balpha}$ with at least  one reverse-aligned element. Let us denote the number of reverse-aligned elements corresponding to $\mathcal{Q}_r$ by $r_o$, where $r_o\geq 1$. Then, the transition probability from $\mathcal{Q}_r$ to the absorbing state $\mathcal{Q}_1$ is $(1/N)^{r_o}(1-1/N)^{N-r_o}$. We can therefore conclude that the absorbing state can be reached from all the states $\mathcal{Q}_r$ so that the Markov chain is an \emph{absorbing Markov chain} (see \cite[Def. 11.1]{Grinstead1997Introduction-to}). It then follows from \cite[Th. 11.3]{Grinstead1997Introduction-to} that the system moves to the absorbing state $\mathcal{Q}_1$ w.p. $\!1$ as the length of the training phase 
becomes large. Since $\mathcal{Q}_1$ corresponds to the beamforming configuration, we can conclude that the iterative algorithm eventually converges to the beamforming configuration.
\end{itemize}

We emphasize that due to our simplifying assumptions, compared to \cite{Mudumbai2006Distributed-Tra}, of the fading coefficients and the signals being real-valued, we were able to prove convergence of the iterative algorithm to the beamforming configuration in a  simple fashion using basic results from Markov chain theory. The proof above, however, does not reveal anything about the rate of convergence of the algorithm.  It seems difficult to obtain results on the actual convergence rate because the state-transition probabilities in the Markov chain, implicitly, depend on the actual realizations of the fading coefficients $h^j$. However, interesting insights into the behavior of the  convergence rate, as a function of $N$, can be obtained by considering convergence in expectation according to 
 \ba
 \label{SISOInterferenceProbabilisticConvergence}
 \mathbb{E}\big[\hspace{-1mm}\sum_{\mathcal{S}^j\in\,\mathcal{G}}\! h^j\hat{\alpha}^j[t]\big]&=
\mathbb{E}_{h^j}\hspace{-0.6mm}\bigg[\mathbb{E}_{\{\alpha^j[l]\}_{l=0}^t|h^j}\big[\hspace{-1mm}\sum_{\mathcal{S}^j\in\,\mathcal{G}}\!h^j\hat{\alpha}^j[t]\big]\bigg] \notag\\
&\stackrel{t\rightarrow k_oN}{\text{\huge{$\longrightarrow$}}} {\mathbb{E}}_{h^j}\hspace{-0.6mm}\bigg[\sum_{\mathcal{S}^j\in\,\mathcal{G}}|h^j|\bigg].
\ea
In particular, we shall show that convergence in the sense of \eqref{SISOInterferenceProbabilisticConvergence} can be obtained if the length of the training phase $T_{tr}$ scales linearly in $N$, i.e., $k_o$ is independent of $N$. To be precise, we shall establish convergence to the beamforming configuration up to a certain level in the sense that we will allow for a (small) fraction of the sources to be reverse-aligned. The corresponding concept and the associated convergence proof are provided next. 

\subsection{Convergence to $\epsilon$-level and convergence proof}
In the following, we shall be interested in convergence of the iterative algorithm in the sense of \eqref{SISOInterferenceProbabilisticConvergence} up to a certain level. Concretely, we shall allow that, on average, a (small) fraction $\epsilon$ of the sources in a given group is reverse-aligned, which trivially implies that $(1-\epsilon)$ is the fraction of sources that are aligned, on average. Throughout the paper, we shall assume that $\epsilon$ is independent of $N$ and\footnote{Strictly speaking, for a given $N$, this requires that $\epsilon$ be an integer multiple of $1/N$.} $\epsilon N\in\mathbb{Z}^{+}$. Denoting the sets of aligned and reverse-aligned sources after the update \eqref{AuxiliaryParameterUpdateRule} at the end of the $t_{th}$ frame as $\mathcal{A}[t]$ and $\overline{\mathcal{A}}[t]$, respectively, we have
\ba
\label{EqualityCanBeInequality}
\sum_{\mathcal{S}^j\in\,\mathcal{G}}\!h^j\hat{\alpha}^j[t] &= \sum_{\mathcal{S}^j\in\, \mathcal{A}[t]}\!|h^j|-\sum_{\mathcal{S}^j\in\, \overline{\mathcal{A}}[t]}\!|h^j|.
\ea
We say that convergence in the sense of \eqref{SISOInterferenceProbabilisticConvergence} up to an $\epsilon$-level has been achieved if 
\ba
\label{InequalityRatherThanEquality}
&{\mathbb{E}}_{h^j}\bigg[\mathbb{E}_{\{\alpha^j[l]\}_{l=0}^t|h^j}\bigg[\sum_{\mathcal{S}^j\in\,\mathcal{G}}\!h^j\hat{\alpha}^j[t]\bigg]\bigg]\notag\\
&= {\mathbb{E}}_{h^j}\bigg[ \mathbb{E}_{\{\alpha^j[l]\}_{l=0}^t|h^j}\bigg[
\sum_{\mathcal{S}^j\in\,\mathcal{A}[t]}\!|h^j|-\sum_{\mathcal{S}^j\in\,\overline{\mathcal{A}}[t]}\!|h^j|\bigg]\bigg]\\
\label{InequalityRatherThanEquality2}
&\geq N(1-2\epsilon)\Ex{|h^j|}.
\ea

We can now formalize our convergence result as follows.
\begin{theorem}
For any $\epsilon_o>0$ and large $N$, the number of iterations required in the second step of the distributed beamforming algorithm in Section~\ref{sec:SingleAntennaSchemeDetails} to achieve \emph{convergence to $\epsilon_o$-level} is at most $k_oN$, where $k_o$ is a constant independent of $N$.
\end{theorem}

\emph{Proof:} We start by defining  
\ba
\label{SISOInterferenceAllertonDefinitionSNew}
\mathcal{S}^{\text{new}}[t]&=\sum_{\mathcal{S}^j\in\,\mathcal{G}}\!h^j\hat{\alpha}^j[t] - \sum_{\mathcal{S}^j\in\,\mathcal{G}}\!h^j\hat{\alpha}^j[t-1] 
\ea
so that $S^{\text{new}}[t]\geq 0,$ for all $t$. With this definition, we have
\ba
&{\mathbb{E}}_{h^j}\hspace{-0.8mm}\bigg[{\mathbb{E}}_{\{\alpha^j[l]\}_{l=0}^{k_oN-1}|h^j}\hspace{-0.4mm}\bigg[\sum_{\mathcal{S}^j\in\,\mathcal{G}}\!h^j\tilde{\alpha}^j\bigg]\bigg]\notag\\
&= \hspace{-0.4mm}\sum_{t=1}^{k_oN-1}\hspace{-1.8mm}{\mathbb{E}}_{h^j}\hspace{-0.6mm}\big[{\mathbb{E}}_{\{\alpha^j[l]\}_{l=0}^t|h^j}\big[S^{\text{new}}[t]\big]\big]+ {\mathbb{E}}_{h^j}\big[\hspace{-1mm}\sum_{\mathcal{S}^j\in\,\mathcal{G}}\!h^j\big]\\
\label{SISOInterferenceLowerBoundConvergence}
&= \hspace{-0.4mm}\sum_{t=1}^{k_oN-1}\hspace{-1.8mm}{\mathbb{E}}_{h^j}\hspace{-0.6mm}\big[{\mathbb{E}}_{\{\alpha^j[l]\}_{l=0}^t|h^j}\big[S^{\text{new}}[t] \big]\big].
\ea
Let there be $(1-\epsilon)N$ and $\epsilon N$ sources in $\mathcal{A}[t-1]$ and $\overline{\mathcal{A}}[t-1]$, respectively, with $0<\epsilon_o<\epsilon<1$. Moreover, let $q_{a}[t]$ and $\overline{q}_{a}[t]$ denote the number of sources in $\mathcal{A}[t-1]$ and $\overline{\mathcal{A}}[t-1]$, respectively, that alter their beamforming weights from $1$ to $-1$ or vice-versa according to \eqref{RandomPerturbation} at the beginning of the $t_{th}$ frame. Then, we get
\ba
&\mathbb{E}_{h^j}\big[\mathbb{E}_{\{\alpha^j[l]\}_{l=0}^t|h^j}[S^{\text{new}}[t] ]\big]\notag\\
&=P(q_a[t]=0)\ \mathbb{E}_{h^j}\big[\mathbb{E}_{\{\alpha^j[l]\}_{l=0}^t|h^j}[S^{\text{new}}[t] \big| q_a[t]=0]\big]\notag\\
&\quad+P(q_a[t]> 0)\ \mathbb{E}_{h^j}\big[\mathbb{E}_{\{\alpha^j[l]\}_{l=0}^t|h^j}[S^{\text{new}}[t]\big| q_a[t]>0 ]\big]\notag\\
&\geq P(q_a[t]=0)\ \mathbb{E}_{h^j}\big[\mathbb{E}_{\{\alpha^j[l]\}_{l=0}^t|h^j}[S^{\text{new}}[t] \big| q_a[t]=0]\big]\notag\\
&=P(q_a[t]=0)\sum_{s=0}^{\epsilon N}P(\overline{q}_a[t]=s)\notag\\
\label{SISOInterferenceLowerBoundConvergenceFirst}
&\quad\cdot\mathbb{E}_{h^j}\big[\mathbb{E}_{\{\alpha^j[l]\}_{l=0}^t|h^j}[S^{\text{new}}[t] \big| q_a[t]=0,\overline{q}_{a}[t]=s]\big].
\ea
We next show that the expected value inside the summation in \eqref{SISOInterferenceLowerBoundConvergenceFirst} can be lower-bounded by $c_os$, where $c_o$ is a constant independent of $N$. To this end, we start by noting that $q_a[t]=0,\overline{q}_a[t]=s$ corresponds to the case where precisely $s$ sources move from $\overline{\mathcal{A}}[t-1]$ to $\mathcal{A}[t]$ and none of the sources moves from $\mathcal{A}[t-1]$ to $\overline{\mathcal{A}}[t]$.  We denote the sets of these $s$ sources as $\mathcal{R}$, and note that since the $s$ sources in $\mathcal{R}$ are chosen from the $\epsilon N$ sources in $\overline{\mathcal{A}}[t-1]$, there are precisely ${{\epsilon N}\choose s}$ possible choices for $\mathcal{R}$, with the corresponding sets denoted as $\mathcal{R}_1,\mathcal{R}_2,\ldots,\mathcal{R}_{{\epsilon N}\choose s}$. Furthermore, each source in $\overline{\mathcal{A}}[t-1]$ alters its beamforming weight independently and with equal probability, and hence, each of the ${{\epsilon N}\choose s}$ choices is equally likely. From \eqref{SISOInterferenceAllertonDefinitionSNew} it therefore follows that for a given set $\mathcal{R}_i$, we have
\ba
\mathbb{E}_{\{\alpha^j[l]\}_{l=0}^t|h^j}\big[\mathcal{S}^{\text{new}}[t]\big|\mathcal{R}_i\big]=2\vspace{-0.5mm}\sum_{\mathcal{S}^j\in\,\mathcal{R}_i}|h^j|
\ea
which implies
\ba
&\mathbb{E}_{\{\alpha^j[l]\}_{l=0}^t|h^j}[S^{\text{new}}[t] \big| q_a[t]=0,\overline{q}_{a}[t]=s]\notag\\
&\qquad =\frac{1}{{{\epsilon N}\choose s} }
\sum_{\mathcal{R}_i\in\{\mathcal{R}_1,\ldots,\mathcal{R}_{{\epsilon N}\choose s}\}} \mathbb{E}_{\{\alpha^j[l]\}_{l=0}^t|h^j}\big[\mathcal{S}^{\text{new}}[t]\big|\mathcal{R}_i\big]\notag\\
&\qquad =\frac{2}{{{\epsilon N}\choose s} }
\sum_{\mathcal{R}_i\in\{\mathcal{R}_1,\ldots,\mathcal{R}_{{\epsilon N}\choose s}\}} 
\sum_{\mathcal{S}^j\in\,\mathcal{R}_i}|h^j|\notag\\
&\qquad \stackrel{(a)}{=}\frac{2}{{{\epsilon N}\choose s} }\sum_{\mathcal{S}^j\in\,\overline{\mathcal{A}}[t-1]} {{\epsilon N-1}\choose {s-1}}|h^j|\notag\\
&\qquad=\frac{2s}{\epsilon N}\sum_{\mathcal{S}^j\in\,\overline{\mathcal{A}}[t-1]} |h^j|\notag
\ea
where Step $(a)$ is a result of the fact that each source in $\overline{\mathcal{A}}[t-1]$ is present in precisely ${{\epsilon N-1}\choose {s-1}}$ of the sets $\mathcal{R}_i$. We therefore get
\ba
&\mathbb{E}_{h^j}\big[\mathbb{E}_{\{\alpha^j[l]\}_{l=0}^t|h^j}[S^{\text{new}}[t]\big| q_a[t]=0,\overline{q}_{a}[t]=s ]\big]\notag\\
&\qquad\qquad =   \frac{2s}{\epsilon N}\ \mathbb{E}_{h^j}\bigg[\sum_{\mathcal{S}^j\in\,\overline{\mathcal{A}}[t-1]} |h^j|\bigg]  \notag\\
&\qquad\qquad \! \stackrel{(a)}{=}2s\ \mathbb{E}_{h^j}\big[|h^j|\big| \mathcal{S}^j\in\overline{\mathcal{A}}[t-1]\big]\notag\\
\label{UseTheNeatTrick}
&\qquad\qquad=2s\int_x x f_{|h^j|\big|\overline{\mathcal{A}}[t-1]}(x)\,\mathrm{d}x
\ea
where Step $(a)$ follows from $|\overline{\mathcal{A}}[t-1]|=\epsilon N$ and the fact that the $h^j$ are identically distributed. 
Next, we need to show that the integral on the right hand side (RHS) of \eqref{UseTheNeatTrick} can be lower-bounded by a constant independent of $N$. To this end, we start by noting that
\ba
f_{|h^j|}(x)&=P(\overline{\mathcal{A}}[t-1])f_{|h^j|\big|\overline{\mathcal{A}}[t-1]}(x)\notag\\
&\qquad+P(\mathcal{A}[t-1])f_{|h^j|\big|\mathcal{A}[t-1]}(x)\notag\\
&\geq P(\overline{\mathcal{A}}[t-1])f_{|h^j|\big|\overline{\mathcal{A}}[t-1]}(x)\notag\\
&=\epsilon\  f_{|h^j|\big|\overline{\mathcal{A}}[t-1]}(x)\notag
\ea
which, using $\epsilon>\epsilon_o$, implies
\ba
\label{SISOInterferenceConvergenceConstraint}
f_{|h^j|\big|\overline{\mathcal{A}}[t-1]}(x)&\leq \frac{f_{|h^j|}(x)}{\epsilon_o}.
\ea
We can then argue that finding the minimum of $\int_x xf(x)\,\mathrm{d}x$ over the class of functions $f(x)$ that satisfy
\ba
&f(x)= 0,\qquad x<0,\notag\\
&f(x)\geq 0,\qquad x\,\ge\,0,\notag\\
&f(x)\leq\frac{f_{|h^j|}(x)}{\epsilon_o}\quad\text{and}\notag\\
\label{MinimizationConstraintsSISOInterference}
&\int_x  f(x)\,\mathrm{d}x=1
\ea
guarantees that this minimum is a lower bound on $\int_x x f_{|h^j|\big|\overline{\mathcal{A}}[t-1]}(x)\,\mathrm{d}x$ as $ f_{|h^j|\big|\overline{\mathcal{A}}[t-1]}(x)$ is a member of this class of functions. Concretely, we want to determine
\ba
\min_{f(\cdot)} \int_x x f(x)\,\mathrm{d}x
\ea
where the minimization is under the constraints \eqref{MinimizationConstraintsSISOInterference}. This minimization problem can be solved as follows. We start by setting $x_o= \mathcal{Q}^{-1}\hspace{-1mm}\left(\frac{1-\epsilon_o}{2}\right)$, where $\mathcal{Q}(v)=\int_{v}^\infty\frac{1}{\sqrt{2\pi}}e^{-\frac{u^2}{2}}\,\mathrm{d}u$, so that using $f_{|h^j|}(x)=\sqrt{\frac{2}{\pi}}e^{-\frac{x^2}{2}},\ x\,\ge\,0$, and $f_{|h^j|}(x)=0,\ x<0$, we have
\ba
\int_{0}^{x_o} \frac{f_{|h^j|}(x)}{\epsilon_o}\,\mathrm{d}x=1.\notag
\ea
Next, we note that
\ba
&\int_{0}^{\infty}\!x f(x)\,\mathrm{d}x =\int_{0}^{x_o}\!x f(x)\,\mathrm{d}x+\int_{x_o}^{\infty}\!xf(x)\,\mathrm{d}x\notag\\
&\geq \int_{0}^{x_o}\!xf(x)\,\mathrm{d}x+x_o\int_{x_o}^{\infty}\!f(x)\,\mathrm{d}x\notag\\
&=\int_{0}^{x_o}\!x f(x)\,\mathrm{d}x+x_o\bigg[1-\int_{0}^{x_o}\!f(x)\,\mathrm{d}x\bigg] \notag\\
&=\int_{0}^{x_o}\!x f(x)\,\mathrm{d}x+x_o\bigg[\int_{0}^{x_o}\!\frac{f_{|h_j|}(x)}{\epsilon_o}\,\mathrm{d}x-\int_{0}^{x_o}\!f(x)\,\mathrm{d}x\bigg]\notag\\
&=x_o\int_{0}^{x_o}\!\frac{f_{|h_j|}(x)}{\epsilon_o}\,\mathrm{d}x-\int_{0}^{x_o}\!(x_o-x) f(x)\,\mathrm{d}x\notag\\
&=\int_{0}^{x_o} \!x \ \frac{f_{|h_j|}(x)}{\epsilon_o}\,\mathrm{d}x+\int_{0}^{x_o}\!(x_o-x) \frac{f_{|h_j|}(x)}{\epsilon_o}\,\mathrm{d}x\notag\\
&\qquad-\int_{0}^{x_o}\!(x_o-x) f(x)\,\mathrm{d}x\notag\\
\label{AllertonResultOfMinimizationTrick}
&=\int_{0}^{x_o}\!\!\!x\ \frac{f_{|h_j|}(x)}{\epsilon_o}\,\mathrm{d}x+\int_{0}^{x_o}\!\!(x_o-x) \bigg(\frac{f_{|h_j|}(x)}{\epsilon_o}-f(x)\bigg) \mathrm{d}x. 
\ea
Since the second term in the last line of \eqref{AllertonResultOfMinimizationTrick} is positive due to the constraint $f(x)\leq (1/\epsilon_o)f_{|h^j|}(x)$, it follows that setting 
\ba
\label{AllertonDistributionThatMinimizes}
f(x)&=
\begin{cases}
\frac{f_{|h^j|}(x)}{\epsilon_o},&\qquad 0\leq x\leq x_o\\
0,&\qquad\text{otherwise}
\end{cases}
\ea
yields the desired lower bound. Further, corresponding to $f(x)$ in \eqref{AllertonDistributionThatMinimizes}, we get
\ba
\label{AllertonMinimizedValueCo}
\int_x x f(x)\,\mathrm{d}x&=\left(\frac{1-e^{-\frac{\left(\mathcal{Q}^{-1}\left(\frac{1-\epsilon_o}{2}\right)\right)^2 }{2}}}{\epsilon_o}\right)\sqrt{\frac{2}{\pi}}\define\frac{c_o}{2}.
\ea
Substituting \eqref{AllertonMinimizedValueCo} into \eqref{UseTheNeatTrick} and the result thereof into  \eqref{SISOInterferenceLowerBoundConvergenceFirst}, we obtain 
\ba
&\mathbb{E}_{h^j}\big[\mathbb{E}_{\{\alpha^j[l]\}_{l=0}^{t}|h^j}[S^{\text{new}}[t] ]\big]\notag\\
&\geq c_o P(q_a[t]=0)\sum_{s=0}^{\epsilon N}sP(\overline{q}_a[t]=s)\\
&=c_o\left(1-\frac{1}{N}\right)^{(1-\epsilon)N}\\
&\qquad \cdot \sum_{s=1}^{\epsilon N}s {{\epsilon N}\choose s} \left(\frac{1}{N}\right)^s\left(1-\frac{1}{N}\right)^{\epsilon N-s}\\
&=c_o\left(1-\frac{1}{N}\right)^{N}\sum_{s=1}^{\epsilon N}s {{\epsilon N}\choose s} \left(\frac{\frac{1}{N}}{1-\frac{1}{N}}\right)^{s}\\
&=c_o\left(1-\frac{1}{N}\right)^{N} \sum_{s=1}^{\epsilon N}s{{\epsilon N}\choose s} \left(\frac{1}{N- 1}\right)^{s}\\
&=c_o\left(1-\frac{1}{N}\right)^{N} \frac{1}{N-1}\ \epsilon N\hspace{-1mm}\left(1+\frac{1}{N- 1}\right)^{\epsilon N-1}\\
&\quad\big(\text{since $\sum_{k=1}^K{K\choose k}k x^{k-1}=K(1+x)^{K-1}$}\big)\notag\\
&=c_o\left(1-\frac{1}{N}\right)^{N} \frac{\epsilon N}{N-1}\left(\frac{N}{N-1}\right)^{\epsilon N-1}\\
&=c_o\ \epsilon  \left(1-\frac{1}{N}\right)^{(1-\epsilon)N} \\
&\approx c_o\ \epsilon\ e^{-(1-\epsilon)}\qquad\text{(for large $N$)} \\
\label{NumberOfNewSourcesAlignedLowerBound}
&>  c_o\ \epsilon_o\ e^{-(1-\epsilon_o)}\qquad\text{(since $\epsilon>\epsilon_o)$}.
\ea
Substituting \eqref{NumberOfNewSourcesAlignedLowerBound} into \eqref{SISOInterferenceLowerBoundConvergence} finally yields 
\ba
{\mathbb{E}}_{h^j}\hspace{-0.6mm}\bigg[{\mathbb{E}}_{\{\alpha^j[l]\}_{l=0}^{k_oN-1}|h^j}\hspace{-0.6mm}\bigg[\sum_{\mathcal{S}^j\in\,\mathcal{G}}h^j\tilde{\alpha}^j\bigg]\bigg]
&> \sum_{t=1}^{k_oN-1}\frac{c_o\epsilon_o} {e^{(1-\epsilon_o)}}\notag\\
&=(k_oN-1)\frac{c_o\epsilon_o}{ e^{(1-\epsilon_o)}}\notag\\
&\approx k_oN\frac{c_o\epsilon_o}{ e^{(1-\epsilon_o)}}
\ea
which upon setting 
\ba
k_o=\frac{(1-2\epsilon_o)e^{1-\epsilon_o}}{c_o\epsilon_o}\Ex{|h^j|}\notag
\ea
and noting that $k_o$ does not depend on $N$ establishes the desired result. $\Box$
\begin{figure*}
\centerline{\subfigure[\text{N=100}]{\includegraphics[width=80mm]{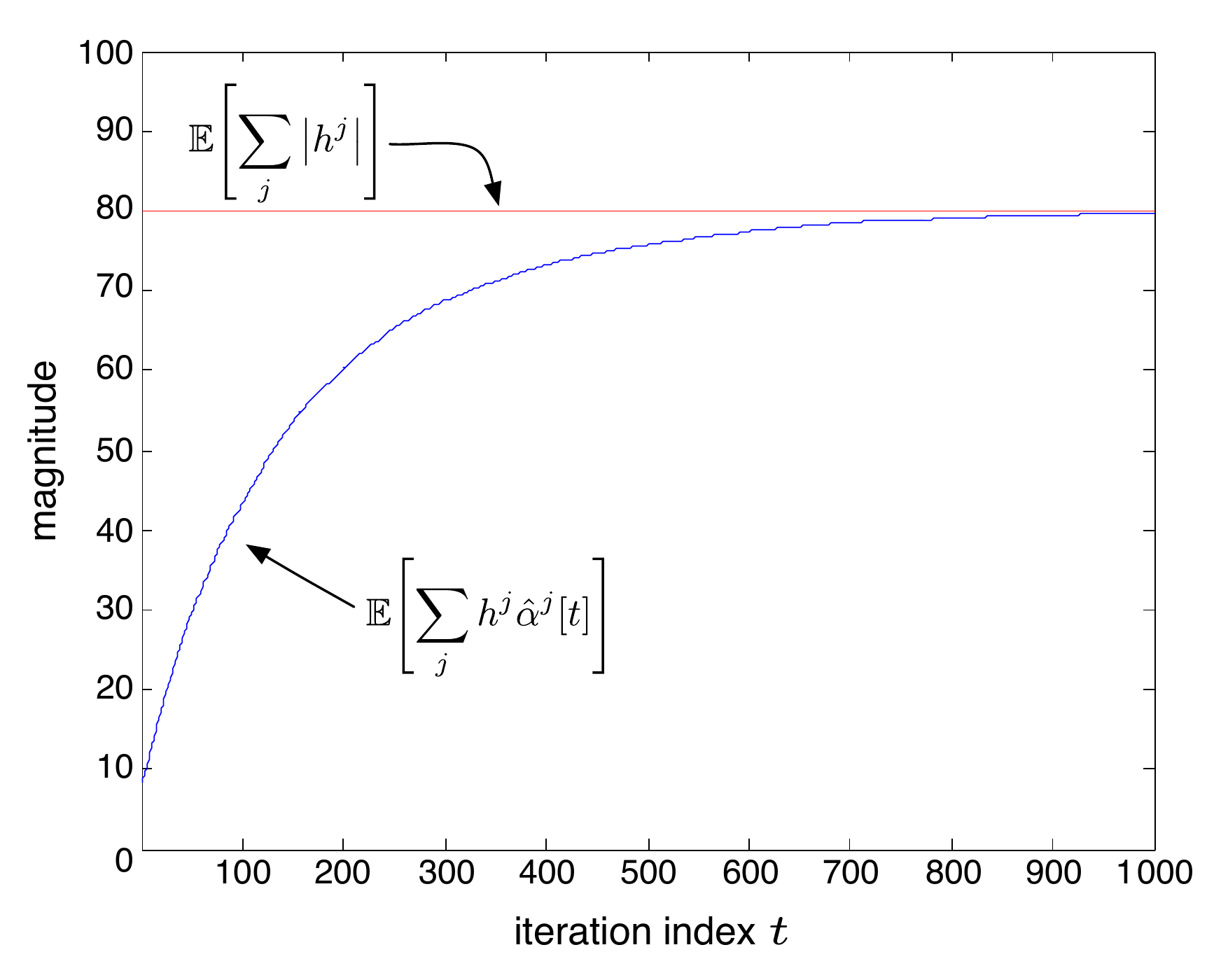}
\label{SimulationN100K1000T50Label}}
\hfil
\subfigure[\text{N=500}]{\includegraphics[width=80mm]{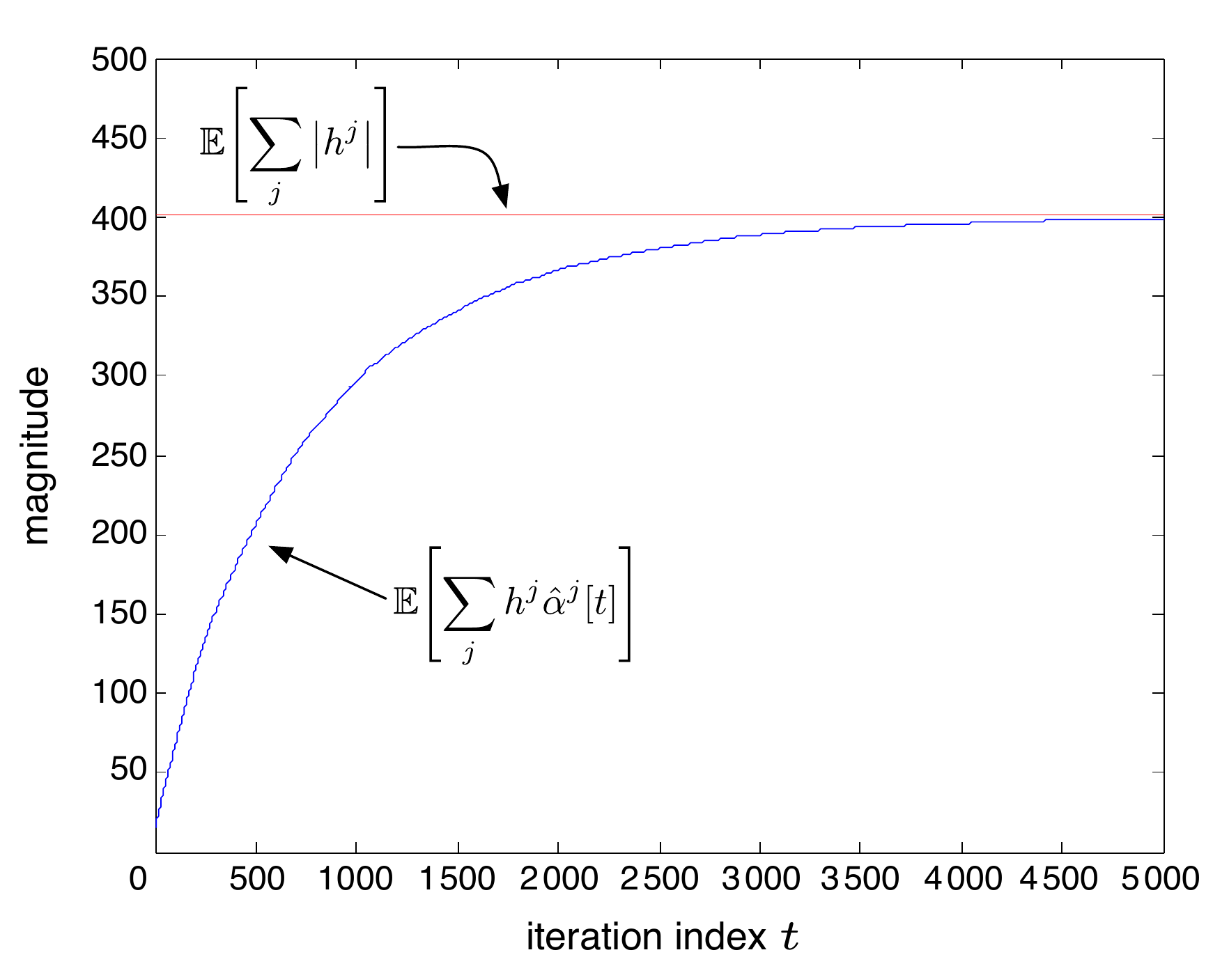}
\label{SimulationN500K5000T50Label}}}
\centerline{\subfigure[\text{N=1000}]{\includegraphics[width=80mm]{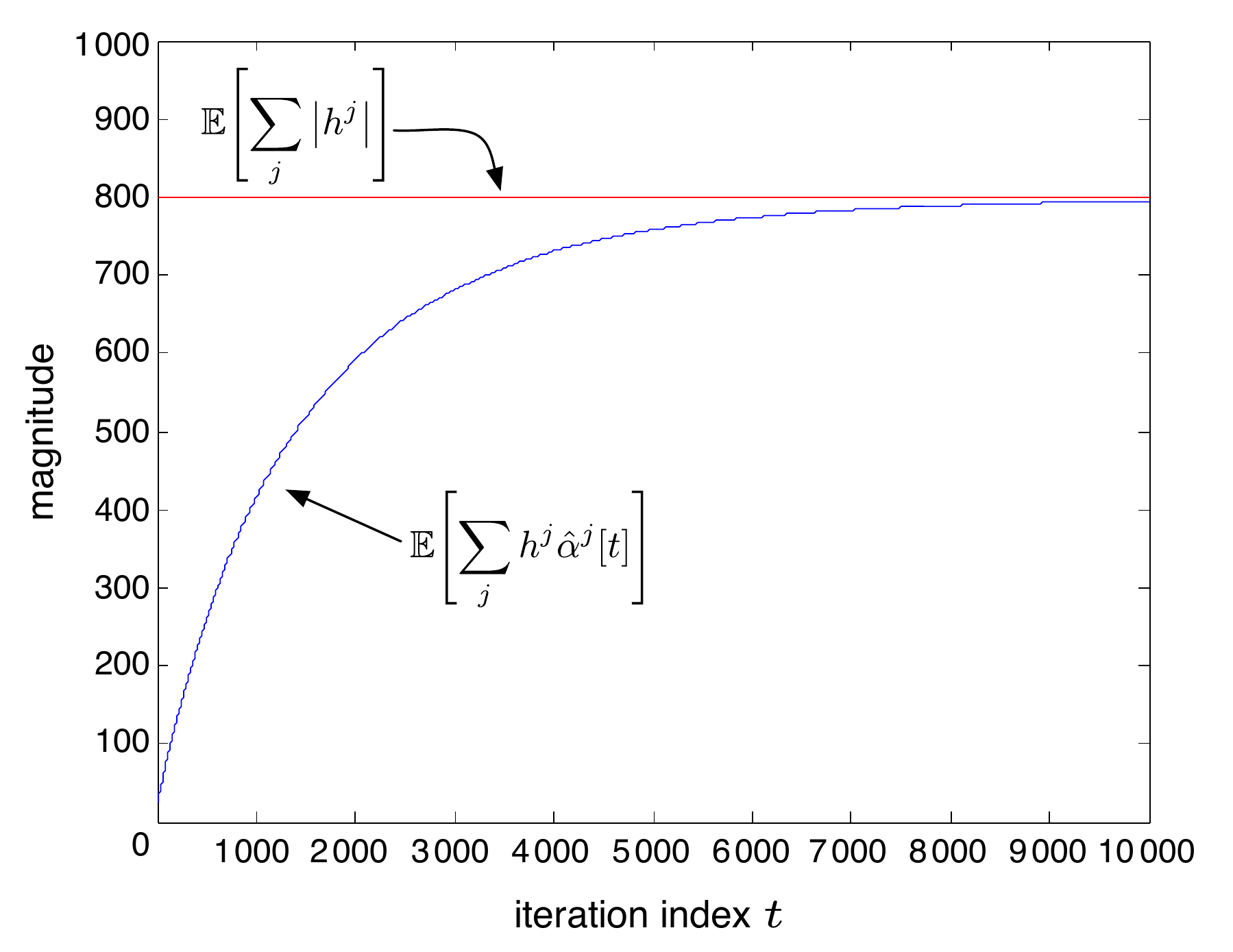}
\label{SimulationN1000K10000T50Label}}
\hfil
\subfigure[\text{N=5000}]{\includegraphics[width=80mm]{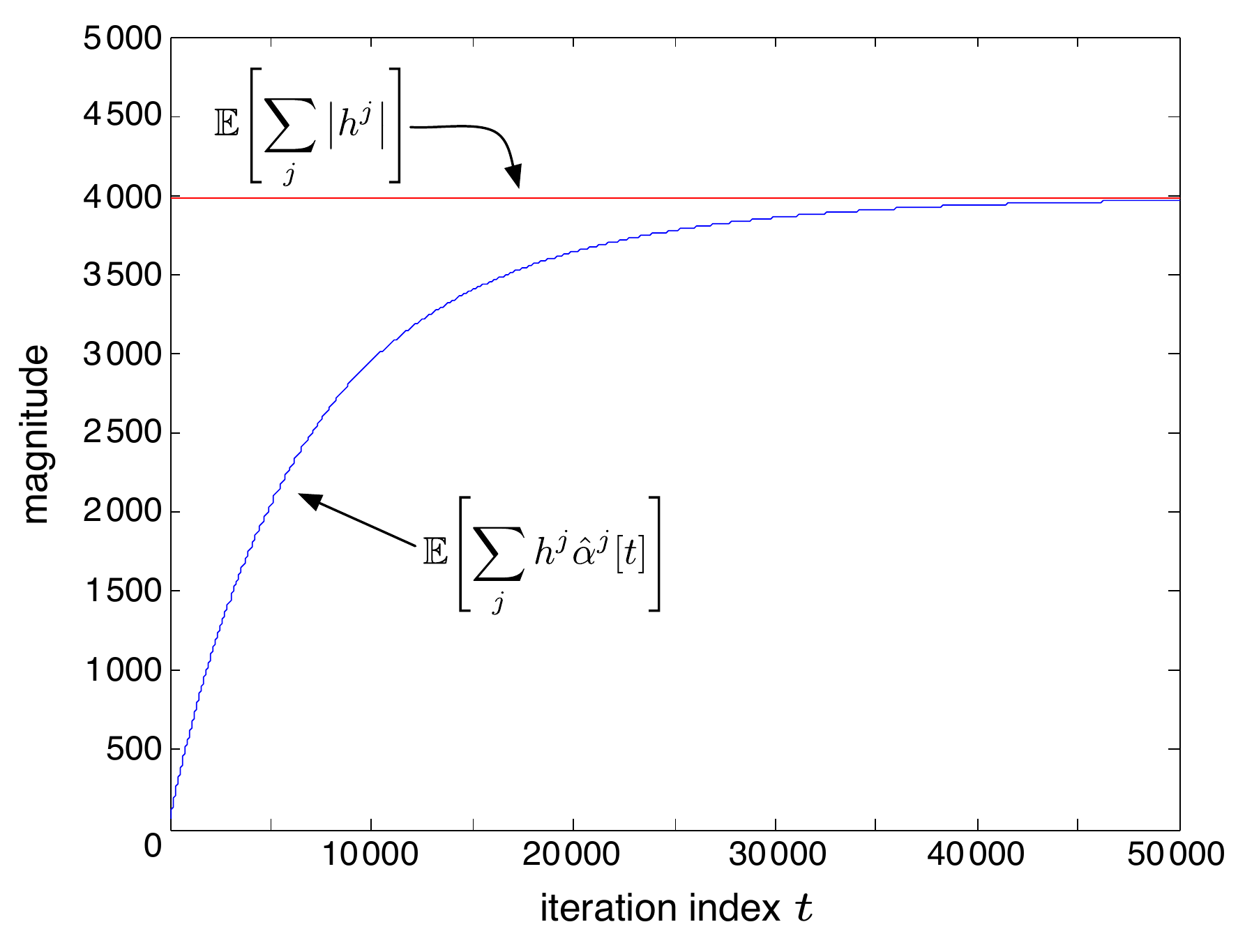}
\label{SimulationN5000K50000T50Label}}}
\caption{Convergence of the iterative distributed beamforming algorithm.}
\label{SISOInterferenceConvergenceSimulation}
\end{figure*}

\subsection{\it{Numerical Results}}

Next, we present simulation results related to the convergence behavior of the iterative distributed beamforming algorithm described in Section~\ref{sec:SingleAntennaSchemeDetails}. In particular, Fig.\ref{SISOInterferenceConvergenceSimulation} shows how $\Ex{\sum_{\mathcal{S}^j\in\,\mathcal{G}}h^j\hat{\alpha}^j[t]}$ (averaged over $50$ realizations of the fading coefficients $h^j$) evolves as a function of $t$ for $N=100,500,1000$ and $5000$, respectively. In all four cases, convergence occurs within approximately $10N$ \hspace{-0.5mm}iterations, thereby corroborating the fact that the convergence time is linear in $N$.


\section{Achievability of Multiplexing Gain and Crystallization}
\label{sec:SingleAntennaProofOfAchievability}
The aim of this section is to show that performing data transmission, i.e., spatial multiplexing, based on the beamforming weights $\tilde{\alpha}_i^j$ obtained from the iterative distributed beamforming algorithm described in Section~\ref{sec:SingleAntennaSchemeDetails} results in a spatial multiplexing gain of $M$ and a per-stream (distributed) array gain proportional to $N$. Moreover, using the framework introduced in\cite{Morgenshtern2006Crystallization}, we prove that the network ``crystallizes'', i.e., the individual fading links $\mathcal{G}_i\rightarrow\mathcal{D}_i$ converge to non-fading links as $N\rightarrow\infty$. Finally, we  quantify the impact of $M$ and $\epsilon_o$ on the crystallization rate, i.e., the rate, as a function of $N$, at which the fading links  $\mathcal{G}_i\rightarrow\mathcal{D}_i$ converge to non-fading links.

Consider the group $\mathcal{G}_i$  and let $\tilde{\mathcal{A}}_i$ and $\overline{\tilde{\mathcal{A}}}_i$ denote the set of aligned and reverse-aligned sources, respectively, corresponding to the beamforming weights $\tilde{\alpha}_i^j$, so that $|\tilde{\mathcal{A}}_i|=(1-\epsilon_o)N$ and $|\overline{\tilde{\mathcal{A}}}_i|=\epsilon_oN$. Then, the corresponding input-output relations for the individual links $\mathcal{G}_{i}\rightarrow\mathcal{D}_{i}$ are 
\ba
y_i[n]&=x_i[n]\left(\sum_{\tilde{\mathcal{A}}_i}|h_{i,i}^j |-\sum_{\overline{\tilde{\mathcal{A}}}_i}|h_{i,i}^j |\right) \notag\\
&+\underbrace{\mathop{\sum_{r=1}^M}_{r\neq i}\bigg( \sum_{j=1}^N h_{i,r}^j\tilde{\alpha}_r^j\bigg)x_r[n]}_{\text{interference}}\ +\ w_i[n],\ i=1,2,\ldots,M.\notag
\ea
Note that here we assume that for any given realization of the fading coefficients $h_{i,i}^j$, precisely $(1-\epsilon_0)N$ of the sources are aligned. Convergence of the iterative distributed beamforming algorithm in expectation according to \eqref{InequalityRatherThanEquality2}, however, guarantees only that this is the case on average. We furthermore assume Gaussian codebooks in what follows.
Under the assumption of the destination $\mathcal{D}_i$ having perfect knowledge of its effective channel coefficient $(\sum_{\tilde{\mathcal{A}}_i}|h_{i,i}^j |-\sum_{\overline{\tilde{\mathcal{A}}}_i}|h_{i,i}^j|)$ 
and of the coefficients $\sum_{j=1}^N h_{i,r}^j\tilde{\alpha}_r^j, \mbox{for all}\,\, r\neq i,$ corresponding to the effective  interference channels, the outage probability for the link $\mathcal{G}_i \rightarrow \mathcal{D}_i$ is given by
\ba
&P_i^{\text{out}}(R_i) = P\left(\frac{1}{2}\log(1+\mathrm{SINR}_i)<   R_i \right)\\
&= P\left(\mathrm{SINR}_i<  2^{2R_i}-1  \right)\\
&= P\left(\frac{ {\frac{P}{N}} \big( \sum_{\tilde{\mathcal{A}}_i}|h_{i,i}^j|- \sum_{\overline{\tilde{\mathcal{A}}}_i} |h_{i,i}^j|  \big)^2}{\sum_{r\neq i}\frac{P}{N} \left| \sum_{j}h_{i,r}^j\tilde{\alpha}_r^j\right|^2+N_o} <   2^{2R_i}-1 \right)\\
\label{FirstBound}
&= P\left( \frac{ \big( \sum_{\tilde{\mathcal{A}}_i}|h_{i,i}^j|- \sum_{\overline{\tilde{\mathcal{A}}}_i} |h_{i,i}^j|  \big)^2}{\sum_{r\neq i}\left| \sum_{j}h_{i,r}^j\tilde{\alpha}_r^j\right|^2+\frac{NN_o}{P}} <  2^{2R_i}-1   \right).
\ea
To upper-bound the outage probability, we use the (union) bounding techniques introduced in 
\cite{Morgenshtern2006Crystallization}. Skipping the details, we note that for any positive $k_1$, $k_2$ and $k_3$ such that $k_1>k_2$, the following upper bound holds:
\ba
&P\left(        \frac{ \left( \sum_{\tilde{\mathcal{A}}_i}|h_{i,i}^j| - \sum_{\overline{\tilde{\mathcal{A}}}_i} |h_{i,i}^j|  \right)^2}{\sum_{r\neq i}\left| \sum_{j}h_{i,r}^j\tilde{\alpha}_r^j\right|^2   +\frac{NN_o}{P}} <           \frac{(k_1-k_2)^2}{k_3}\right) \notag\\
\label{FirstBoundForUnionBound}
&\leq P\hspace{-1mm}\left(\left( \sum_{\tilde{\mathcal{A}}_i}|h_{i,i}^j| - \sum_{\overline{\tilde{\mathcal{A}}}_i} |h_{i,i}^j|  \right)^2  <     (k_1-k_2)^2\right) \notag\\
&\qquad+  P\hspace{-1mm}\left( \sum_{r\neq i}\left| \sum_{j}h_{i,r}^j\tilde{\alpha}_r^j\right|^2 +\frac{NN_o}{P}>  k_3 \right)\\
&\leq P\hspace{-1mm}\left(\left( \sum_{\tilde{\mathcal{A}}_i}|h_{i,i}^j| - \sum_{\overline{\tilde{\mathcal{A}}}_i} |h_{i,i}^j|  \right)^2  <     (k_1-k_2)^2\right) \notag\\
&\qquad +  \sum_{r\neq i}P\left( \left| \sum_{j}h_{i,r}^j\tilde{\alpha}_r^j\right|^2 >  \frac{k_3-\frac{NN_o}{P}}{M-1} \right)\\
&\leq P\bigg( \sum_{\tilde{\mathcal{A}}_i}|h_{i,i}^j|  <  k_1 \bigg) +P\bigg( \sum_{\overline{\tilde{\mathcal{A}}}_i} |h_{i,i}^j|   > k_2\bigg)\notag\\
\label{ChangedArgumentToMakeExplicit}
&\qquad +  \sum_{r\neq i}P\hspace{-1mm}\left(\left| \sum_{j}h_{i,r}^j\tilde{\alpha}_r^j\right| >  \sqrt{\frac{k_3-\frac{NN_o}{P}}{M-1}} \right)
\ea
where we implicitly assumed that $M>1$ and $k_3\geq (NN_o)/P$. Next, we use the fact, noted previously at the end of Section~III, that 
$h_{i,r}^j\tilde{\alpha}_r^j\sim h_{i,r}^j$, for $r\neq i$. Consequently, we have
\ba
&P\hspace{-1mm}\left(   \frac{ \big( \sum_{\tilde{\mathcal{A}}_i}|h_{i,i}^j| - \sum_{\overline{\tilde{\mathcal{A}}}_i} |h_{i,i}^j|  \big)^2}{\sum_{r\neq i}\big| \sum_{j}h_{i,r}^j\tilde{\alpha}_r^j\big|^2   +\frac{NN_o}{P}} <         \frac{(k_1-k_2)^2}{k_3}\right) \notag\\
&\leq P\bigg( \sum_{\tilde{\mathcal{A}}_i}|h_{i,i}^j|  <  k_1 \bigg) +P\bigg( \sum_{\overline{\tilde{\mathcal{A}}}_i} |h_{i,i}^j|   > k_2\bigg)\notag\\
\label{AlmostBeforeSubstitutingFromAppendix}
&\qquad+  (M-1) P\hspace{-1mm}\left(\left| \sum_{j}h_{i,r}^j \right| >  \sqrt{\frac{k_3-\frac{NN_o}{P}}{M-1}} \right).
\ea
\addtolength{\textheight}{-4cm}   
Each of the three terms in \eqref{AlmostBeforeSubstitutingFromAppendix} can be upper-bounded  
using large deviations bounds\cite{Shwartz1995Large-Deviation}. In particular, employing 
\ba
P\left( \sum_{i=1}^{S}|h_{i}| < k \right) &\leq \left(\frac{ek}{S}\sqrt{\frac{2}{\pi}}\right)^S\notag\\
P\left( \sum_{i=1}^{S}|h_{i}| > k \right)&\leq\left(2e^{-\frac{k^2}{2S^2}}\right)^S\notag\\
P\left(\sum_{i=1}^S h_i \geq k\right) &\approx e^{-\frac{k^2}{2S}}\notag
\ea
in \eqref{AlmostBeforeSubstitutingFromAppendix}, we get
\ba
\label{SubstituteFromAppendix}
P_i^{\text{out}}(R_i)&\leq \left(\frac{ek_1}{(1-\epsilon_o)N}\sqrt{\frac{2}{\pi}}\right)^{(1-\epsilon_o)N}+
\left(  2e^{-\frac{k_2^2}{2(\epsilon_oN)^2}}\right)^{\epsilon_oN}\notag\\
&\qquad+2(M-1)e^{-\frac{k_3-\frac{NN_o}{P}}{2N(M-1)}}.
\ea

The key to obtaining meaningful upper bounds from \eqref{SubstituteFromAppendix} lies in a judicious  choice of the constants $k_1,k_2$ and $k_3$ ensuring that, for $R_i=(1/2)\log(1+(k_1-k_2)^2/k_3)$, $P^{\text{out}}_i(R_i)\rightarrow 0$ as $N\rightarrow \infty$ while satisfying the conditions $k_1>k_2$ and $k_3\geq (NN_o)/P$. Since $\Ex{\sum_{\tilde{\mathcal{A}}_i}|h_{i,i}^j|}\,\propto\,N$ and $\text{VAR}\hspace{-1mm}\left[\sum_{\tilde{\mathcal{A}}_i} |h_{i,i}^j|\right] \propto N$, motivated by the above large deviations bounds, it is sensible to set
 \ba
 \label{SISOInterferenceChoiceK1}
 k_1&=\frac{(1-\epsilon_o)(N-\sqrt{N})}{e}\sqrt{\frac{\pi}{2}}.
 \ea
Similarly, since $\sum_{\overline{\tilde{\mathcal{A}}}_i} |h_{i,i}^j| $ deviates around a mean value proportional to $N$ with a variance proportional to $N$ and $|\sum_{j}h_{i,r}^j|$ deviates around a mean value proportional to $\sqrt{N}$ 
with a variance proportional to $N$, again motivated by the above large deviations bounds, we set
\ba
\label{SISOInterferenceChoiceK2K3}
k_2&=\sqrt{2(1+\ln 2)}(\epsilon_oN+\sqrt{N})\notag\\
\text{and}\quad k_3&=(M-1)N^{1+\delta}+\frac{NN_o}{P} 
\ea
with the constant $\delta>0$. Note that the condition $k_1\,>\,k_2$ implies that $\epsilon_{0}\,<\,\frac{1}{1+e\sqrt{\frac{4}{\pi}(1+\ln 2)}}=0.2419$. With the
above choices for the parameters $k_1,k_2$ and $k_3$,  in the limit $N\rightarrow\infty$, we get
\ba
 2^{2R_i-1}&= \frac{(k_1-k_2)^2}{k_3} \notag\\
 &= \frac{N^{1-\delta}}{(M-1)}\left(\frac{1-\epsilon_o}{e}\sqrt{\frac{\pi}{2}}-\sqrt{2(1+\ln 2)} \epsilon_o\right)^2\notag\\
 \label{SISOInterferenceAllertonDefineC1}
& \define c_1N^{1-\delta} 
\ea
so that $R_i=(1/2)\log\left( 1+c_1N^{1-\delta}\right)$. Substituting \eqref{SISOInterferenceChoiceK1} and \eqref{SISOInterferenceChoiceK2K3} into \eqref{SubstituteFromAppendix}, in the large-$N$ limit, we finally obtain
\ba
&P_i^{\text{out}}\left(\frac{1}{2}\log\left( 1+c_1N^{1-\delta}\right)\right)\notag\\
&\leq \left(1-\frac{1}{\sqrt{N}}\right)^N + e^{-\epsilon_o N-2\sqrt{N}}+  2(M-1) e^{-\frac{N^{\delta}}{2}}\notag\\
\label{SISOInterferenceAlmostDone}
&= e^{-\sqrt{N}}+e^{-\epsilon_o N-2\sqrt{N}}+ 2(M-1)e^{-\frac{N^{\delta}}{2}}.
\ea

We can therefore conclude that $P_i^{\text{out}}(R_i)\rightarrow 0$ as $N\rightarrow\infty$ for any rate $R_i\leq(1/2)\log(1+c_1N)$ (recall that $\delta$ can be arbitrarily small). Since this holds true for all groups $\mathcal{G}_i$, we can choose $R_i=(1/2)\log(1+c_iN)$ with $c_i<c_1$, for all $i$, and get $P_i^{\text{out}}(R_i)\rightarrow 0$, for all $i$, as $N\rightarrow\infty$, which implies full spatial multiplexing gain of $M$, a per-stream array gain proportional to $N$, and convergence of each of the links $\mathcal{G}_i\rightarrow\mathcal{D}_i$ to a non-fading link. In summary, in the language of \cite{Morgenshtern2006Crystallization}, we can conclude that the network ``crystallizes'' as $N\rightarrow\infty$. The third term on the RHS of \eqref{SISOInterferenceAlmostDone} nicely reflects the impact of interference on the crystallization rate. Specifically, for $\delta<1/2$, this term dominates the decay rate as a function of $N$. A smaller $\delta$ corresponds, through $R_i=(1/2)\log(1+c_1N^{1-\delta})$, to higher data rates, but results in a reduced crystallization rate. In the single-user case, i.e., for $M=1$, the third term equals zero reflecting the absence of interference. We can therefore conclude that the crystallization rate in the presence of interference is significantly smaller than in the single-user case  $M=1$. 
Finally, regarding the proportionality constant $c_1$, it can be  observed that the smaller $\epsilon_o$ (i.e., the smaller the fraction of reverse-aligned sources) the larger $c_1$, and hence the larger the individual rates $R_i=(1/2)\log(1+c_iN)$ still guaranteeing crystallization. On the other hand, for smaller $\epsilon_o$ the second term on the RHS of \eqref{SISOInterferenceAlmostDone} becomes larger, again reflecting that a higher data rate comes at the cost of increased outage probability.  

\addtolength{\textheight}{-3cm}   
\bibliographystyle{IEEEtran}
\bibliography{IEEEabrv,MuxGainZFBeamforming}
\end{document}